# High Energy & High Luminosity γγ Colliders

Emanuela Barzi*

*Fermi National Accelerator Laboratory, Batavia, IL 60510, USA, and Ohio State University, Columbus, OH 43210, USA*

Barry C. Barish

*California Institute of Technology, Pasadena, CA 91125, USA, and U.C. Riverside, Riverside, CA 92521, USA*

William A. Barletta

*Massachusetts Institute of Technology, Cambridge, MA 02139, USA*

Simone Di Mitri

*Elettra Sincrotrone Trieste, 34149 Basovizza, Trieste, ITALY, and University of Trieste, Trieste 34100, Italy*

Ilya F. Ginzburg

*Sobolev Inst. of Mathematics of SB RAS + Novosibirsk State University, Novosibirsk Oblast, Russia, 630090*

ABSTRACT

With the best of modern standard lasers, high-energy γγ colliders from electron beams of E ≥ 250 GeV are only possible at the expense of photon luminosity, i.e. 10 times lower than for photon colliders at c.m. energies below 0.5 TeV. For existing state-of-the art lasers, an optimistic upper energy limit for x=4.8 is an electron beam of less than 250 GeV. This Snowmass21 Contributed Paper shows how Free Electron Lasers (FEL) pave the way for High Energy & High Luminosity γγ colliders. We present and assess a conceptual design study of a FEL with wavelength of 2.4 μm and an x-factor in the range of 2 to 40, to maximize the luminosity of a gamma gamma collider as second interaction region of 0.5 TeV to 10 TeV c.m. e+e- colliders.

## 1. EXECUTIVE SUMMARY AND RECOMMENDATIONS

### 1.1. Summary

With the best of modern standard lasers, high-energy γγ colliders from electron beams of E ≥ 250 GeV are possible at the expense of photon luminosity, or 1% of the geometric e+e- luminosity, i.e. 10 times lower than for photon colliders at c.m. energies below 0.5 TeV. This is because the higher energy γ rays annihilate in the Breit-Wheeler process, $\gamma\gamma_0 \rightarrow$ e+e- and in the Bethe-Heitler disrupting process, $e\gamma \rightarrow e\, e+e-$. These processes occur when the invariant quantity $x=12.3\, E\,(\text{TeV})/\lambda_0\,(\mu m)$ exceeds 4.8, where $E$ is the energy of each e+ or e-beam, and $\lambda_0$ is the wavelength of the incoming photon. Instead, for x<4.8, the luminosity of the γγ collisions can be as high as 43% of the geometric e+e- luminosity, which translates to 10% of the geometric luminosity at the desired γγ c.m. energy.

We show how a single Free Electron Laser (FEL) design meets the specifications to enable γγ colliders as second interaction regions of e+e- colliders over the energy range of 0.5 TeV to 10 TeV c.m. without sacrificing γγ luminosity. The same electron beams and accelerators of the original e+e- collider are used for two identical high gain Self-Amplified Spontaneous Emission (SASE) FELs. At the appropriate energy required in the FEL design, i.e. 2.3 GeV, every other bunch from each beam is diverted to each FEL line where a helical undulator produces circularly polarized 0.5 eV light with 0.1 - 1 Joules per pulse. The remaining bunches continue down the Linac until reaching their nominal energy, and colliding with geometric luminosity of 1-6 × $10^{34}$cm$^2$/s. The central FEL wavelength of 2.4 μm, obtained with either standard warm magnet or superconducting technology for the undulator, and an x-factor in the range of 2 to 40, maximize the luminosity of the γγ collider as second interaction region of a 0.5–10 TeV c.m. electron-positron collider.

*barzi@fnal.gov

This FEL increases the expected γ intensity by a factor of 10 for electron beams of up to 0.5 TeV, by a factor of 6 for electron beams of up to 1.5 TeV, and by a factor of 3 for electron beams of up to 5 TeV. This translates to a 10-fold increase in the luminosity of γγ colliders as second interaction regions of 0.5 TeV to 1 TeV c.m. e+e- colliders, a factor of 6 for a 3 TeV c.m. e+e- collider, and a factor of 3 for a 10 TeV c.m. e+e- collider. This FEL concept therefore paves the way for High Energy & High Luminosity γγ colliders.

## 1.2. Recommendations

The FEL concept proposed in this white paper produces a 10 factor gain in the luminosity of γγ colliders as second interaction region of e+e- colliders up to at least 1 TeV c.m.. We therefore recommend that a γγ collider be considered a natural part of all linear collider proposals. The richer and complementary physics produced by such an integrated machine at only the added cost of the FEL undulators and the electron/positron transfer lines would make the case for an ILC collider stronger and for an ILC more likely to be built.

## 2. INTRODUCTION

γγ colliders have a long history [1]-[3], are a rich multidisciplinary field, and since the late 1980s have been considered a natural part of all linear collider proposals. For e+e- colliders which use their beams only once, high energy photons with a brightness comparable to that of the electron beam are generated through Compton backscattering of laser light focused onto the incoming electron bunch just before the interaction point [4]. The fractional energy transfer from the electron beam to the photons goes from 46% at the 45 GeV beam energy typical of Z factories, up to 83% at 250 GeV beam energy. The photons with the maximum energy are produced in the direction of the electron beam. One laser pulse per bunch transfers energy from both electron beams into photon beams which would collide in the interaction point (IP). eγ collisions also occur between each photon beam and its opposite electron beam. When the scattering is designed to occur at a minimal distance from the IP, the photon spectrum is more monochromatic. The center of mass (c.m.) energy in the electron-laser photon interaction determines the maximum possible fractional energy transfer, and the polarizations of the electron and laser beam control the produced photon energy and polarization distribution. By using circularly polarized laser beams it is possible to produce highly circularly polarized photon beams at the maximum energy.

The physics case for γγ colliders starts from the lowest energy options all the way up to the multi-TeV concepts. The γγ approach on its own eliminates the challenges of building a powerful positron source and subsequent fast damping ring. At the lowest end of the energy spectrum, a γγ collider with 12 GeV c.m. energy based on the 17.5 GeV superconducting linac of the European XFEL [5] would make uniquely possible the study of γγ physics in the bb̄ region. The discovery of a Higgs-like boson at 125 GeV in 2012 renewed interest in a dedicated electron linac-based

Higgs factory from γγ collisions. The proposed SAPPHiRE [6] would use recirculating electron linac technology to create 80 GeV electron beams and produce γγ collisions at 125 GeV c.m. Similarly, HFiTT [7] proposed to use the Fermilab's Tevatron tunnel to accelerate up to 80 GeV two electron beams for γγ production at 125 GeV. An X-ray FEL-based γγ collider Higgs factory with two colliding 63 GeV electron beams is proposed at SLAC using Cool Copper Collider technology [8].

The fundamental design criterion for a photon collider laser is that it should supply enough laser light that every electron bunch has sufficient (>60%) electrons undergo a primary Compton backscatter. For an individual laser pulse, the pulse width should be at ~ps scale, and the pulse energy from a fraction to several J. The specifications of the laser itself depend on the choice of e+e- collider parameters, which include the time structure of the electron beam, the electron beam energy, and the electron bunch length. Over the whole spectrum of proposed e+e- colliders, the average laser power per beam spans from 6.4 kW for [8], to 33 kW for the 500 GeV ILC stage, to 89 kW for a 500 GeV CLIC, to 239 kW for HFiTT [7], and up to 1000 kW for SAPPHiRE [6]; the laser energy per bunch train from 5 J for SAPPHiRE and HFiTT, to 27 J for [8], to 1770 J for a 500 GeV CLIC, and up to 6600 J for the 500 GeV ILC stage.

With the best of modern standard lasers, high-energy γγ colliders from electron beams of E ≥ 250GeV are possible at the expense of photon luminosity, i.e. 10 times lower than for photon colliders at c.m. energies below 0.5 TeV [9]. Studies of the physics of γγ and eγ collisions using a 500 GeV ILC stage include Higgs boson searches in the direct production mode γγ ––> H° and single gauge boson production. In the case of a 1 TeV linear collider to produce colliding photons, the unique physics that could be performed includes: hadron physics through the study of the photon structure Wγ; purer high energy QCD processes, such as diffraction, total cross section; multiple production of gauge bosons; the detailed structure of the electroweak theory with 0.1% accuracy through processes such as γγ ––> WW, eγ ––> νW, and γγ ––> ZZ; strongly interacting scalars beyond the Standard Model thanks to the machine high monochromaticity. We herein show how Free Electron Lasers (FEL) can be designed to pave the way for High Energy & High Luminosity γγ colliders. We present and assess a conceptual design study of the laser and/or Free Electron Laser that would be required for γγ colliders from 0.5 TeV to 10 TeV c.m. energy.

## 3. γγ COLLIDER CONCEPTS

A very clear paper to explain γγ concepts is "GAMMA-GAMMA COLLIDERS" by Kwang-Je Kim & Andrew Sessler [10], and we are therefore going to use it extensively in this Section. The γγ colliders contemplated in our study are to be added to an electron-positron collider of various c.m. energies. This is a way to obtain two machines that are complementary in their physics capabilities at a much reduced capital cost. A schematic of such a system is shown in Fig. 1. The γ rays are produced by Compton back-scattering of low energy photons $\gamma_0$ when hitting a high energy electron or positron beam. The collision point is also called conversion point. The γ rays obtained in this manner travel in the same direction as the original electron beam with an energy distribution that peaks near 83% of the electron energy. As Fig. 2 illustrates, this peak in the γ ray energy occurs when the helicity of the electrons is opposite to that of the incoming photons $\gamma_0$. In order for the two γ rays to collide with each other in their own interaction region, another low energy photon beam $\gamma_0$ is shot against the positron beam of the e+e- collider. The γ rays intensity depends on the intensity of the incoming photon beam $\gamma_0$. To exploit the e+e- luminosity, from the cross section of the Compton process and the laser pulse area, one finds that $10^9$ photons are needed to hit each electron and produce one back-scattered $\gamma$ ray, or $10^{19}$ photons for an electron bunch of $10^{10}$ electrons.

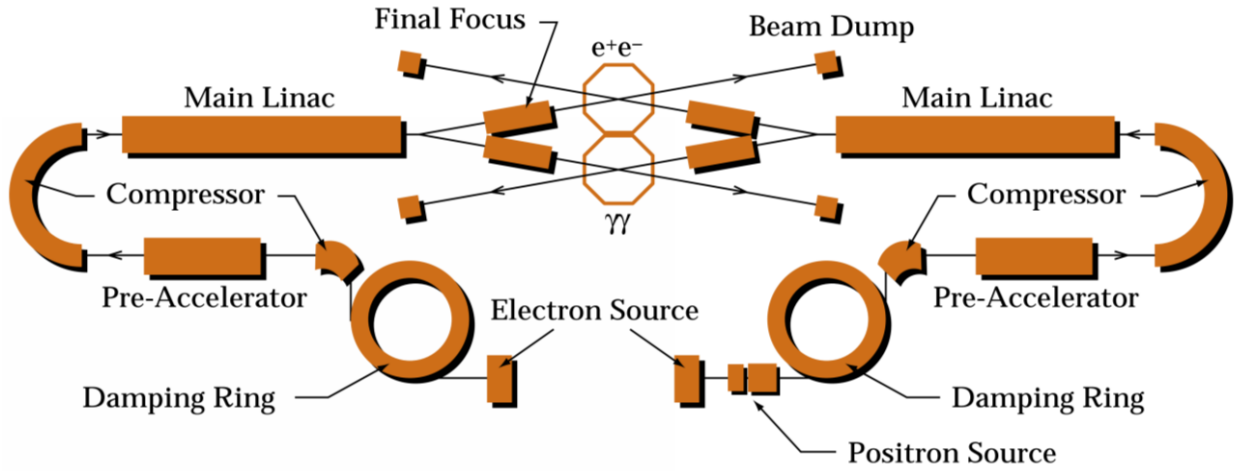

Figure 1: Schematic of an e+e- collider which includes a second interaction region for $\gamma\gamma$ collisions [10].

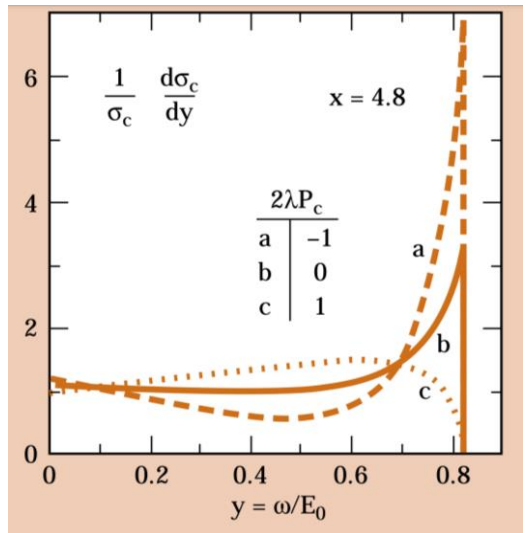

Figure 2: Spectrum of the Compton scattered photons for different combinations of helicities [10].

The highest possible $\gamma\gamma$ luminosity is obtained for an invariant quantity x=12.3 $E$ (TeV)/$\lambda_0$ (μm), where $E$ is the energy of each e+ or e-beam, and $\lambda_0$ is the wavelength of the incoming photon, smaller than 4.8. This is because for x>4.8 the higher energy $\gamma$ rays annihilate in the Breit-Wheeler process, $\gamma\gamma_0 \rightarrow$ e+e-. For x<4.8, the luminosity of the $\gamma\gamma$ collisions can be as high as 43% of the geometric e+e- luminosity, which translates to 10% of the geometric luminosity at the desired $\gamma\gamma$ c.m. energy. For x>4.8, it has been shown that with appropriate parameters optimization, the luminosity of the $\gamma\gamma$ collisions at the desired $\gamma\gamma$ c.m. energy is no more than 1% of the geometric e+e- luminosity [9].

The specifications of the laser to produce the incoming photon $\gamma_0$ include a required wavelength $\lambda$ of ~1 µm, a ~0.1-10 J pulse of ps length, same high repetition rate as that of the electron beam, and variable polarization. An overview of specifications for envisioned e+e- colliders is shown in Table 1. Table 2 details their repetition rates. For existing state-of-the art lasers, an optimistic upper energy limit for x=4.8 is an electron beam of less than 250 GeV. There are yet no lasers that satisfy the requirements to increase the produced $\gamma$ energy at or above 250 GeV without sacrificing $\gamma$ ray intensity. This is detailed in Appendix A.

Table 1. Summary of approximate parameters of an electron-positron collider, assuming round beams from two identical C-band linacs, at the final energy of 0.25 TeV, and flat beams inspired from CLIC design parameters in the presence of crossing angle and crab angle, for the final beam energy range 0.25 – 1.5 TeV [11].

| e+e- collider parameter | Round beam | Flat beam | Units |
|---|---|---|---|
| Electron/positron beam energy | 0.25 | 0.25 – 1.5 | TeV |
| Bunch charge | 1 | 1 | nC |
| Bunch repetition rate | 1 | 6.6 – 16.2[*] | kHz |
| Particles per bunch | $0.65 \cdot 10^{10}$ | $0.65 \cdot 10^{10}$ | |
| Normalized rms emittance (x,y) | 0.5 | 0.8, 0.03 | µm rad |
| Betatron function at IP (x,y) | 1, 1 | 7, 0.1 | mm |
| Rms beam size at ee-IP (x,y) | 30 | ~100, ~2 | nm |
| Geometric peak e+e- luminosity | $\sim 1 \times 10^{34}$ | $1 - 6 \times 10^{34}$ | cm$^{-2}$sec$^{-1}$ |
| Site length | ~20 | ~10–20 | km |

[*]See Table 2.

Table 2. Repetition rates for four e+e- colliders [11].

| e+e- collider | c.m. energy, TeV | Beam energy, TeV | Train rep. rate, Hz | No. bunches/train | Tot. rep. rate, kHz |
|---|---|---|---|---|---|
| NLC | 0.5 | 0.25 | 180 | 90 | 16.2 |
| ILC | 0.5 | 0.25 | 5 | 1312 | 6.6 |
| C$^3$ | 0.55 | 0.275 | 120 | 75 | 9.0 |
| CLIC | 3.0 | 1.5 | 50 | 312 | 15.6 |

## 4. FEL DESIGN

FELs have several attractive advantages as a source of laser photons for a photon collider. Picosecond and sub-picosecond bunches are inherently simpler because the laser pulse width is inherently the same as the electron bunch width, and it is possible to produce the required photon wavelength of ~ 1 µm. Unlike atomic lasers, FELs provide some tuneability in the central wavelength through either variable gap (magnetic field) of the undulator, or electron beam energy, or both, e.g. in the range ~ 0.5 – 2.5 µm. This guarantees a maximum luminosity of the $\gamma\gamma$ collider in accordance to a variable c.m. energy of the lepton collider, e.g. in the range 0.5 – 2 TeV. The FEL also allows tunability in pulse duration, for instance for the optimization of the Compton back-scattering conversion efficiency in the presence of a non-zero crossing angle.

However, FELs for γγ colliders have to be specifically designed to generate very high pulse energies in helical undulators to provide the correct circular photon polarization. ANL for instance has extensive experience with superconducting bifilar helical coil configurations [12]. The design of superconducting FEL undulators would include, in addition to the magnets, also phase shifters, focusing magnets and precise electron beam position monitors. The FELs considered are of the high gain Self-Amplified Spontaneous Emission (SASE) type.

### 4.1. Electron Accelerator for FEL

In our case, the γγ collide at a second interaction point of an e+e- collider of given c.m. energy $2E$. We have studied two possible scenarios.

**(1).** In the first scenario, the same electron and positron accelerators of the original e+e- collider are used for the FEL. This guarantees that the key specification of same high repetition rate as that of the electron/positron beam be met. At the appropriate energy required in the FEL design, i.e. 2.3 GeV, every other bunch from each beam is diverted to each FEL line where a helical undulator produces circular polarized 0.5 eV light with 0.1-1 Joules per pulse. The remaining bunches continue down the Linac until reaching their nominal energy $E$, and colliding with geometric luminosity of 1-6 $\times 10^{34}$ cm$^2$/s. At the Compton conversion point, the electrons of energy $E$ collide with the low energy photons from the opposing FEL line to produce photons of maximum energy $E \cdot x/(1+x)$. The two identical FEL lines provide the necessary circular polarization at the resonance wavelength of 2.4 μm (0.5 eV) through either conventional or superconducting undulators. Compton backscattering on 0.25-5 TeV electron and positron beams (see specs below) can produce photons in the energy range 0.1-4.9 TeV.

No issues preserving the emittance of a high brightness 2.3 GeV electron/positron beam over the several kilometers to the FEL are foreseen. By designing a transport line with small angles, coherent synchrotron radiation from the bends can be controlled, and with it the electron/positron beam emittance growth. This method has the added advantage that -- compared to a design with separate 2.3 GeV and 5 TeV Linacs -- it should be easier to maintain synchronization between the FEL photon beam and the 5 TeV e- beam at the Compton conversion point.

**(2).** In a second scenario, a separate FEL option of 1.2 μm and fixed 1 kHz repetition rate is also envisaged for a hypothetical 0.5 TeV c.m. e+e- collider, either with round beam and CW mode or flat beam in "burst" mode, and whose specs are shown in the first column of Table 1. This FEL is based on C-band linac technology and has a reduced footprint.

### 4.2. FEL Configuration for 1 TeV c.m. e+e- Collider

For the 1 TeV c.m. energy of two head-on colliding electron-positron beams, a 1.8 T-peak magnetic field of the undulator is considered. An undulator period of 15 cm corresponds to a relatively large undulator parameter K=25. We choose x=4. This determines a central wavelength of 0.17 TeV for the Compton backscattered photons at the Compton edge, which in turn identifies the optimum 2.384 μm laser wavelength for the Compton source. The aforementioned undulator parameters determine a resonant electron beam energy of 2.3 GeV to drive the high gain SASE FEL.

In order to get an FEL energy per pulse of 0.1 J (to be compared with 0.05 J in [3]), an electron peak current of 6 kA and a bunch charge of 1 nC are envisaged. This current level is routinely achieved in present, far more energetic, x-ray FELs, such as for instance SACLA (Japan), which operates with a current of up to 10 kA. The bunch duration at the undulator entrance is therefore estimated to be 50 fs rms (an almost flat-top current profile after magnetic compression is assumed). The required FEL peak power at saturation results ~1 TW, which implies an FEL parameter, or normalized energy bandwidth, larger than 4%. By assuming an electron beam average transverse size along the undulator smaller than 35 μm, we can ensure a FEL parameter larger than 5%, which is perfectly consistent with the aforementioned specifications. As a matter of fact, the electron beam size at the energy of 2.3 GeV is consistent with, for instance, a conservative normalized transverse emittance of 0.7 mm mrad, and average betatron function of 7 m along the undulator. Such beam sizes are consistent with both the round beam and the flat beam parameters in Table 1.

The FEL saturation length, estimated to be as large as 21-times the FEL gain length of 0.15 m, is 3 m in the 1-D approximation of FEL dynamics. This extremely small value is due to the very high electron beam energy resonant at a rather long central wavelength. The FEL cooperation length is approximately one-tenth of the electron bunch duration. Considering some reduction of the output FEL power due to radiation slippage and 3-D effects (electron beam non-zero emittance, relative energy spread, radiation diffraction), the saturation length will not exceed 20 m. Such undulator length would allow further increase of the FEL pulse energy by means of undulator tapering, potentially rising the pulse energy up to 1 J.

The condition $x < 4.8$ avoids the annihilation process of Compton backscattered photons into electron-positron couples. The Bethe-Heitler disrupting process, $e\gamma \rightarrow e\,e+e-$, is avoided as long as $x < 8$.

The normalized optical length of the FEL flash is estimated in the range $z = 0.8\text{-}1$ with the proposed parameters, which allows the optimization of the $\gamma\gamma$ luminosity as discussed in [9]. At the same time, the distance between the conversion point (FEL on collider electron beam) and the photon IP is assumed of the order of 50 mm to make the normalized impact parameter $b_n \approx 1$. This way, the Compton backscattered photons contribute more to the luminosity and its spectrum peaks in correspondence of the maximum photon energy [9]. To produce one $\gamma$ photon per electron, the FEL photons transverse area at the conversion point has to match the Thomson cross section times the number of FEL photons per pulse. Because of diffraction, the former is of the order of the FEL wavelength multiplied by the pulse length (which is approximately the electron bunch length), or 36 μm$^2$ in our case. Hence, the number of FEL photons per pulse should be at least $1.4 \times 10^{18}$. At the FEL photon energy of 0.52 eV, this corresponds exactly to 0.1 J, as anticipated in our FEL design. In summary, the proposed set of parameters $z \approx b_n \approx 1$, $x = 4$, together with an FEL intensity $> 3 \times 10^{14}$ W/cm$^2$ at the interaction point, guarantees an optimum $\gamma\gamma$ luminosity in accordance to the calculations in [9], i.e., ~10% of the e+e- luminosity.

### 4.3. FEL Concepts for 0.5 – 10 TeV c.m. e+e- Colliders

The 2.4 μm FEL design described above can be identically applied to c.m. energy of the e+e-collider in the range 0.5–10 TeV. In such case, the Compton backscattered photon energy would span 0.17–4.9 TeV. Despite the x-parameter growing monotonically from 2 to 40 with increasing energy of the colliding electron beam, the FEL intensity would still ensure a γγ luminosity close to the optimum, around 10% of the e+e- luminosity for the lower c.m. energies of the e+e- collider, down to approximately 3% at the highest energy of 10 TeV c.m..

The separate FEL option in the second scenario of a hypothetical 0.5 TeV c.m. e+e- collider with round beam is described below. In this case, the proposed electron accelerator to drive the FEL relies on well-established C-band RF technology (5.721 GHz), with an important industrial support behind it. The proposal extends the C-band solution presently available in the market to CW 1 kHz repetition rate. This implies the production of 1 kHz C-band klystrons or solid state amplifiers. The choice of a C-band accelerating gradient, estimated to be approximately 15 MV/m at 1 kHz based on present state-of-the-art performance of 55 MV/m at 100 Hz, represents an optimum solution in terms of compactness of the linac footprint, and linac reliability (low breakdown rate). The high brightness electron beam produced in a C-band photo-injector gun and accelerated in the C-band linac, eliminates the need of damping rings. The whole linac tunnel for the maximum energy herein considered of 1.5 GeV, is shorter than 150 m. Since the C-band technology is fully compatible with the relaxed specifications for the electron beam brightness at the injector exit and, consequently, at the undulator entrance, the proposed scheme offers a robust and highly reliable linac design. While the proposed electron beam parameters to drive the FEL represent a conservative choice, they offer some important margin of improvement, e.g. in terms of FEL pulse energy and average power, with dedicated R&D, on a time scale of few years from now.

For the 0.5 TeV c.m. e+e- collider, either with round beam and CW mode or flat beam in "burst" mode, by increasing x from 2 to 4, and therefore without substantial impact on the expected γγ luminosity, the FEL wavelength can be lowered to 1.2 μm, as suitably driven by a 1.5 GeV electron linac. A superconducting undulator of period length 10 cm and undulator parameter K=15 would correspond to a peak magnetic field of 1.6 T. Because of the lower electron beam energy, the FEL parameter is reduced to 3%, but still large enough to guarantee 0.3 TW peak power at saturation. The saturation length is shorter than 10 m, and some additional length can be considered to increase the peak power to ~1 TW level through tapering.

A summary of the specifications obtained in the FEL designs is in Table 3. The parameter x, the Compton backscattered maximum photon energy, and the γγ luminosity are shown in Table 4 as function of the e+e- collider energy in the c.m.. A single FEL design meets the specs of e+e- colliders over the whole energy range from 0.5 TeV to 10 TeV c.m.. This means that once installed, the FEL can be used even if the e+e- collider itself undergoes energy upgrades. But more importantly, as shown in Table 4, this FEL increases the expected γ intensity by a factor of 10 for electron beams of up to 0.5 TeV, by a factor of 6 for electron beams of up to 1.5 TeV, and by a factor of 3 for electron beams of up to 5 TeV. This translates to a 10 factor increase in the luminosity of γγ colliders as second interaction regions of 0.5 TeV to 1 TeV c.m. e+e- colliders, a factor of 6 for a 3 TeV c.m. e+e- collider, and a factor of 3 for a 10 TeV c.m. e+e- collider. This FEL concept therefore paves the way for High Energy & High Luminosity γγ colliders.

As a note, the Ti:Sa ANGUS laser at DESY (2.5 J, 0.8 μm) is close to the specs for the 0.5 TeV c.m. case in pulse energy, but only runs at 5 Hz. Other Nd:YAG 1 kHz, 1.1 μm lasers exist, and used to generate 0.3 μm, 2 mJ, 50 fs pulses, but are not yet optimal in wavelength. The powerful atomic lasers presently under development still have to demonstrate reliable operation at >1 kHz repetition rate. Moreover, they cannot be tuned in wavelength, whereas the proposed FEL allows varying the c.m. energy of the e+e- collider by 20%.

Table 3. Summary of approximate FEL parameters for e+e- collider energy between 0.5 and 10 TeV. A shorter FEL wavelength is also considered for the 0.5 TeV case. "CW" in Table stands for continuous wave.

| FEL parameters | 0.5 TeV | 0.5 - 10 TeV | Units |
|---|---|---|---|
| Electron energy | 1.5 | 2.3 | GeV |
| Repetition rate, CW | 1 | 6.6 – 16.2 | kHz |
| Linac length | < 150 | < 200 | m |
| Bunch charge | 1 | 1 | nC |
| Normalized rms emittance | < 0.7 | < 0.7 | μm rad |
| Relative energy spread, rms | < 0.1 | < 0.05 | % |
| Undulator period | 10 | 15 | cm |
| Undulator peak field | 1.6 | 1.8 | T |
| Undulator parameter K | 14.5 | 25 | |
| Undulator length | < 10 | < 20 | m |
| Average betatron functions | 5 – 7 | 5 – 7 | m |
| FEL resonant wavelength | 1.2 | 2.4 | μm |
| FEL pulse energy | ≥ 0.04 | ≥ 0.1 | J |
| FEL pulse duration, rms | 50 | 50 | fs |
| FEL peak power | ≥ 0.4 | ≥ 1 | TW |
| FEL average power | ≥ 40 | ≥ 100 | W |
| FEL intensity | $1\times10^{14}$ | $3\times10^{14}$ | W/cm$^2$ |
| FEL photons/pulse | ~$0.6\times10^{18}$ | ~$1.4\times10^{18}$ | |

Table 4. x-factor, Compton backscattered maximum photon energy and γγ luminosity as function of the e+e- collider c.m. energy.

| γγ collider parameters | 0.5 TeV | 1.0 TeV | 3.0 TeV | 10 TeV | Units |
|---|---|---|---|---|---|
| **x-factor** | 2 (4) | 4 | 12 | 40 | |
| **Max. photon energy** | 0.17 (0.20) | 0.40 | 1.38 | 4.88 | TeV |
| **L$_{\gamma\gamma}$ / L$_{ee}$** | ≤ 10 | ≤ 10 | ≤ 6 | ≤ 3 | % |

## 5. CONCLUSIONS

To produce γγ colliders as second interaction regions of e+e- colliders over the energy range of 0.5 TeV to 10 TeV c.m., at 2.3 GeV every other bunch from each electron/positron beam is diverted to two identical high gain SASE FEL lines, where a helical undulator produces circularly polarized 0.5 eV light with 0.1-1 Joules per pulse in a footprint of approximately 5 x 20 m$^2$ each. The central FEL wavelength of 2.4 μm, obtained with either standard warm magnet or superconducting technology for the undulator, maximizes the luminosity of the γγ with x-factor increasing from 2 to 40 for electron beam energies from 250 GeV to 5 TeV. At least up to 1 TeV c.m., the γγ luminosity reaches approximately 10% of the electron-positron luminosity by virtue of the optimized FEL design, which is a factor of 10 higher than the 1% that had previously been expected.

# APPENDIX A. State-of-the- art Lasers

As seen in this paper, the specifications of the laser to produce the incoming photon $\gamma_0$ include a required wavelength $\lambda$ of ~1 µm, a ~0.1-10 J pulse of ps length, same high repetition rate as that of the electron beam, and variable polarization. We have seen that in the multi-TeV energy region, the higher energy γ rays annihilate in the Breit-Wheeler process $\gamma\gamma_0 \rightarrow$ e+e- for x>4.8, and in the Bethe-Heitler disrupting process e$\gamma \rightarrow$ e e+e- for x>8. Also, the optimum laser wavelength increases proportionally with the energy, and due to nonlinear effects in the Compton scattering, the required laser flash energy increases [3]. In 2001, a photon collider option was conceptually proposed for a 3 TeV CLIC, with a laser parameter x of 6.5 and wavelength of 4.4 µm [3]. However, the following is stated in [3]: "It is not clear at present, which kind of laser would be best suited for a photon collider at this wavelength. Candidates are a gas CO laser, a free electron laser or some solid state laser or a parametric solid state laser." This was still true in 2021 [9], where it is specified that "there is no known laser to use to produce high energies gammas with x<5."

In the following we show a short survey of existing state-ot-the art lasers provided by Jean-Christophe Chanteloup, who is in charge of Coherent Beam Combining (CBC) at the Ecole Polytechnique, running the XCAN project [14]. Ultrashort and high-peak-power laser pulses are needed also in applications beyond fundamental physics, such as metrology or industry. Solid state lasers can provide high-peak powers. However, their repetition rate is limited due to thermal management. On the contrary, fiber-based laser systems exhibit high average power capability, but are limited in terms of high peak power due to mode confinement in the core of the fiber. To overcome this limitation, the concept of Coherent Amplifying Network (CAN) with fiber amplifiers provides a means of reaching both high peak and high average powers. Certain applications such as particle accelerations will require the coherent combining of up to 10000 fibers, and the peak power needed is perhaps barely accessible now.

The table below shows the current performances of the XCAN laser together with a commercial fiber CBC laser (https://www.afs-jena.de/products/high-power-ultrafast-ytterbium/):

|  | w/o post compression cell | | | |
|---|---|---|---|---|
|  | XCAN | afs Yb-300 | | |
| wavelength | 1,03 | 1,03 | 1,03 | m |
| photon energy | 1,93E-25 | 1,93E-25 | 1,93E-25 | J |
| pulse energy | 1,00E-03 | 2,00E-02 | 2,00E-04 | J |
| pulse duration | 4,00E-13 | 3,00E-13 | 3,00E-13 | s |
|  | 400 | 300 | 300 | fs |
| peak power | 2,50E+09 | 6,67E+10 | 6,67E+08 | W |
|  | 0,0025 | 0,0667 | 0,0007 | TW |
| rep rate | 5,00E+05 | 1,00E+05 | 1,00E+07 | Hz |
| average power | 5,00E+02 | 2,00E+03 | 2,00E+03 | W |
|  | 0,5 | 2,0 | 2,0 | kW |
| focusing diameter | 1,00E-03 | 1,00E-03 | 1,00E-03 | cm |
| surface | 7,85E-07 | 7,85E-07 | 7,85E-07 | cm² |
| intensity | 3,18E+15 | 8,49E+16 | 8,49E+14 | W/cm² |
| photons/pulse | 5,18E+21 | 1,04E+23 | 1,04E+21 |  |

The two columns for the AFS commercial laser correspond to both circles on the following image.

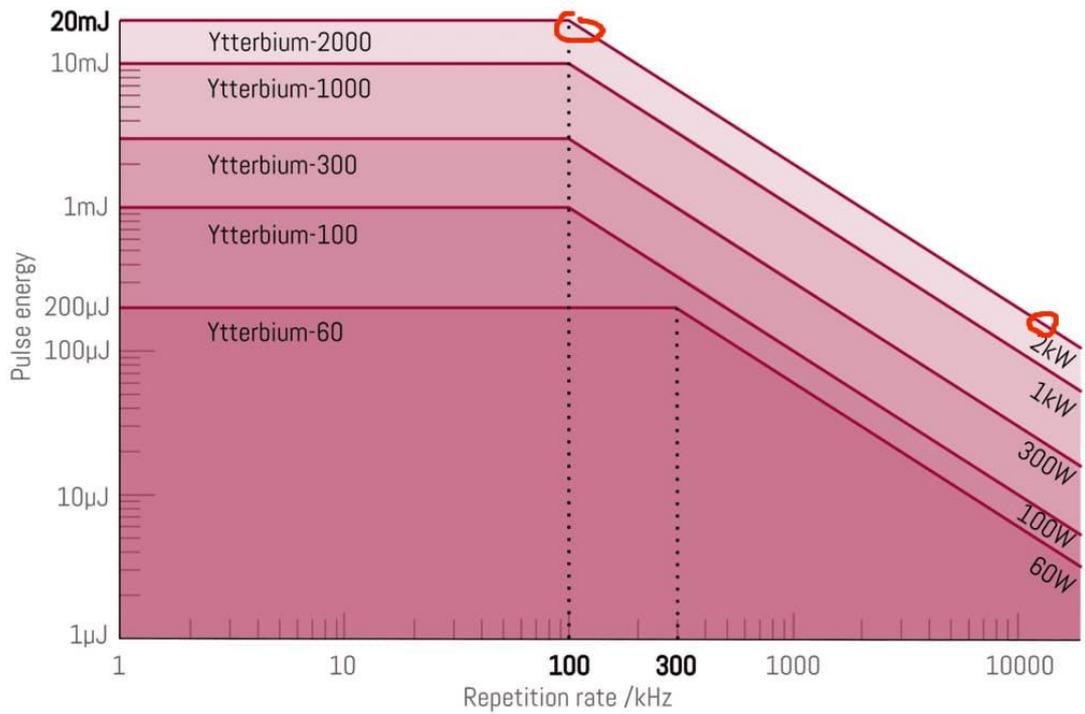

As seen in the table below, the peak powers gain a factor of 10 when operating these lasers with a post compression cell at ~90% efficiency.

|  | | w/ post compression cell | | |
|---|---|---|---|---|
|  | XCAN | | afs Yb-300 | |
| wavelength | 1,03 | 1,03 | 1,03 | m |
| photon energy | 1,93E-25 | 1,93E-25 | 1,93E-25 | J |
| pulse energy | 1,00E-03 | 2,00E-02 | 2,00E-04 | J |
| pulse duration | 4,00E-14 | 3,00E-14 | 3,00E-14 | s |
|  | 40 | 30 | 30 | fs |
| peak power | 2,50E+10 | 6,67E+11 | 6,67E+09 | W |
|  | 0,0250 | 0,6667 | 0,0067 | TW |
| rep rate | 5,00E+05 | 1,00E+05 | 1,00E+07 | Hz |
| average power | 5,00E+02 | 2,00E+03 | 2,00E+03 | W |
|  | 0,5 | 2,0 | 2,0 | kW |
| focusing diameter | 1,00E-03 | 1,00E-03 | 1,00E-03 | cm |
| surface | 7,85E-07 | 7,85E-07 | 7,85E-07 | cm² |
| intensity | 3,18E+16 | 8,49E+17 | 8,49E+15 | W/cm² |
| photons/pulse | 5,18E+21 | 1,04E+23 | 1,04E+21 | |

A second stage of post compression is also feasible to reach sub-10 fs duration but the efficiency drops to 70%, which leads to a 7 times increase in peak power. These lasers are rather compact and efficient (10% wall plug), but operate at a wavelength of 1.03 µm. They are foreseen to be used for electron acceleration in dielectric laser electron acceleration (DLA) or carbon nanotubes (CNT) schemes. Demonstrated DLA was performed with 1.9 µm wavelength laser in silicon structures. With a 1.03 µm laser, silica would be needed. However, structuring a glass substrate at the sub µm level is not as easy on silica.

## Acknowledgments

Emanuela Barzi warmly thanks Tim Barklow, SLAC Linear Accelerator Center, CA, for his expert advice on electron/positron transfer lines, Valery I. Telnov, Novosibirsk State University, Novosibirsk, for useful email discussions, and Jean-Christophe Chanteloup, Ecole Polytechnique, Universite' Paris-Saclay, Palaiseau, France, for his accurate information on state-of-the-art lasers.